\documentclass[pra,twocolumn,nofootinbib,eqsecnum,floatfix,showpacs]{revtex4}
\usepackage{bm} 
\usepackage{amsmath}

\begin{document}

\title{{\bf Born Again}
\thanks{Alberta-Thy-11-09, arXiv:yymm.nnnn [hep-th]}}

\author{Don N. Page}
\email{don@phys.ualberta.ca}

\affiliation{Theoretical Physics Institute\\
Department of Physics, University of Alberta\\
Room 238 CEB, 11322 -- 89 Avenue\\
Edmonton, Alberta, Canada T6G 2G7}

\date{2009 July 23}

\begin{abstract}

A simple proof is given that the probabilities of observations in a large universe are not given directly by Born's rule as the expectation values of projection operators in a global quantum state of the entire universe.  An alternative procedure is proposed for constructing an averaged density matrix for a random small region of the universe and then calculating observational probabilities indirectly by Born's rule as conditional probabilities, conditioned upon the existence of an observation.

\end{abstract}

\pacs{PACS 03.65.Ta, 03.65.Ca, 02.50.Cw, 98.80.Qc, }

\maketitle

\section{Introduction}

In traditional quantum theory, the probabilities of observational results (``observational probabilities'') are given by Born's rule \cite{Born} as the expectation values of projection operators.  This seems to work well in ordinary laboratory settings, where one is considering the observations of a specific observer and knows where he or she is within the quantum state.  However, in cosmology the universe may be so large that there are multiple copies of the observer, and one does not know where the copy is that makes the observation.  In this case there is an additional averaging one must make over the possible locations of the observer, and the resulting probabilities are not given directly by Born's rule from the full quantum state of the universe \cite{HS,typdef,typder,cmwvw,insuff,brd,SH}.
Here a short proof will be given of this fact.  

After this proof of the failure of Born's rule, I shall outline a prescription for replacing it, instead calculating observational probabilities indirectly by forming an average density matrix for each small region and then calculating conditional probabilities for observations by applying Born's rule to this averaged density matrix.

\section{Complete theories of the universe}

Assuming that a complete physical theory of the universe is quantum, I would argue that it should contain at least the following elements:\\
(1) Kinematic variables (wavefunction arguments)\\
(2) Dynamical laws (`Theory of Everything' or TOE)\\
(3) Boundary conditions (specific quantum state)\\
(4) Specification of what has probabilities\\
(5) Probability rules (analogue of Born's rule)\\
(6) Specification of what the probabilities mean

In this paper I shall not consider the questions of what the kinematic variables, dynamical laws, and boundary conditions are.  I shall assume that observational results have probabilities but allow a rather arbitrary choice of what are considered to be observations.

It is then the rules for extracting the probabilities of observations from a rather arbitrarily specified quantum state of the universe that I shall focus on in this Letter.  I shall not address the question of what the probabilities mean, though personally I view them in a rather Everettian way as objective measures for the set of observations with positive probabilities.

In this viewpoint, a goal of science is to produce complete theories $T_i$ that each predict normalized probabilities $P_j(i)$ of observations $O_j$,
\begin{equation}
P_j(i) \equiv P(O_j|T_i)\ \mathrm{with}\  \sum_j P_j(i) = 1.
\label{prob}
\end{equation}

\section{The problem with Born's rule}

Traditional quantum theory uses Born's rule,
\begin{equation}
P_j(i) = \langle \mathbf{P}_j \rangle_i, 
\label{Born}
\end{equation}
where $\mathbf{P}_j$ is the projection operator onto the observational result $O_j$ (or observation $j$, for short), and where $\langle\ldots\rangle_i$ denotes the quantum expectation value, of whatever operator replaces the $\ldots$ inside the angular brackets, in the quantum state $i$ given by the theory $T_i$.  Born's rule works when one knows where the observer is within the quantum state (e.g., in the quantum state of a single laboratory rather than of the universe), so that one has definite orthonormal projection operators.

However, Born's rule does not work in a universe large enough that there may be copies of the observer at different locations, since then one does not know uniquely where the observer is, so that one does not have a definite projection operator $\mathbf{P}_j$ for the observational result.  Then Born's rule is not well defined.

To illustrate this by a toy model, suppose spacetime has $N$ distinct regions labeled by $L$, $1 \leq L \leq N$, but suppose that each observer's observations do not determine which specific region $L$ that observer is in.  Suppose further that one gets a definite projection operator $\mathbf{P}^L_j$ for each observation $j$ within a definite region $L$, obeying $\mathbf{P}^L_j \mathbf{P}^L_k = \delta_{jk}\mathbf{P}^L_j$.  That is, assume that one can neglect the possibility that there are two indistinguishable observers making observations $j$ within any single region $L$, which is what would make it ambiguous which projection operator to use.

However, if the different regions can have copies of the observer which make different observations in different regions, one can have $\mathbf{P}^L_j\mathbf{P}^M_k \neq 0$ for different observations, $j \neq k$, in different regions, $L \neq M$.  That is, the copy of the observer in region $L$ can get the observational result $j$, whereas the copy in region $M$ can get the different result $k$.  Then if the observer tried to use Born's rule to give the probability of observation $j$, it would be ambiguous, since the observer does not know in which region $L$ he or she is.

One might try to use Born's rule with the projection operator to the existence of the observational result in at least one of the regions \cite{HS,typdef,typder,cmwvw,insuff,brd,SH}, $P_j(i) = \langle \mathbf{P}_j^\mathrm{exist} \rangle_i$ with $\mathbf{P}_j^\mathrm{exist} = \mathbf{I} - \prod_L (\mathbf{I} - \mathbf{P}_j^L)$, where here I assume that all the projection operators $\mathbf{P}_j^L$ for the different regions $L$ commute.  But then the resulting probabilities $P_j(i)$ will not be normalized, instead summing to a number greater than unity, since the observational result $O_j$ can occur in one region and $O_k$ within another within the same component of the quantum state.  (That is, the different $\mathbf{P}_j^\mathrm{exist}$'s are not orthogonal.)

This argument shows the basic reason why Born's rule does not work when there may be copies of the observer, as in cosmology with a large enough universe.  Now let me give an explicit proof of this fact.

\section{Proof that Born's rule fails}

To illustrate the problem with Born's rule and prove that one cannot obey
Eq.\ (\ref{prob}) with Born-rule probabilities Eq.\ (\ref{Born}),
let us consider a toy model for a universe in which each component of the quantum state has two regions that can each have either of two observational results, either $O_1$ or $O_2$.  I shall assume that the observational probabilities obey the following principles:

{\it Probability Symmetry Principle\/} (PSP):

If the quantum state is an eigenstate of equal number of observations of
two different observations, then the probabilities of these two
observations are equal.

{\it Prior Rule Principle\/} (PRP):

The rule for extracting probabilities from a quantum state are set down logically prior to the specification of the state; one can consider a different quantum state without changing the rule.

For Born's rule, the PRP says that one should choose the set of orthonormal projection operators logically prior to choosing the quantum state, so that the rule should work with the same projection operators for all quantum states that are then considered.  It is surely not correct to have to know the quantum state in order to choose the projection operators whose expectation values in that state are the probabilities.  In order for the probabilities to depend on the quantum state in some reasonable way, the rule for extracting the probabilities from the state should not be allowed to depend on the quantum state in such an {\it ad hoc} manner.  Here by the Prior Rule Principle, I shall assume that Born's rule is taken with a fixed set of orthonormal projection operators which can then be used for different allowed quantum states.

Now, with the assumptions of the Probability Symmetry Principle and the Prior Rule Principle, consider normalized pure quantum states with $N=2$ regions of the form
\begin{equation}
|\psi\rangle =  b_{12}|12\rangle + b_{21}|21\rangle, 
\label{2-state}
\end{equation}
with the complex amplitudes $b_{12}$ and $b_{21}$ being normalized to obey $|b_{12}|^2 + |b_{21}|^2 = 1$ but otherwise arbitrary.  The component $|12\rangle$ represents the observation $O_1$ in the first region and the observation $O_2$ in the second region; the component $|21\rangle$ represents the observation $O_2$ in the first region and $O_1$ in the second region.

This quantum state is an eigenstate of equal numbers of observations of $O_1$ and of $O_2$, $N_1 = N_2 = 1$ of each, so by the Probability Symmetry Principle the probabilities $P_1(i)$ and $P_2(i)$ should be the same for any theory $T_i$ that gives a quantum state of the form given by Eq.\ (\ref{2-state}) with any allowed complex amplitudes $b_{12}$ and $b_{21}$.  For these equal probabilities to be normalized in this case with only two possible observations, one needs $P_1(i) = P_2(i) = 1/2$ by the PSP.

For Born's rule to give the possibility of both observational probabilities' being nonzero, the orthonormal projection operators should each be of rank one, of the form
\begin{eqnarray}
\mathbf{P}_1 = |\psi_{1}\rangle\langle\psi_{1}|,\ 
\mathbf{P}_2 = |\psi_{2}\rangle\langle\psi_{2}|,
\label{projections}
\end{eqnarray}
where
\begin{eqnarray}
|\psi_{1}\rangle = c_{12}|12\rangle + c_{21}|21\rangle,\ 
|\psi_{2}\rangle = -c^*_{21}|12\rangle + c^*_{12}|21\rangle
\label{projection-states}
\end{eqnarray}
are two orthonormal pure states with complex amplitudes obeying $|c_{12}|^2 + |c_{21}|^2 = 1$.

If one were allowed to choose the projection operators after the quantum state were known, one could choose $c_{12} = (b_{12}+b^*_{21})/\sqrt{2}$ and $c_{21} = (b_{21}-b^*_{12})/\sqrt{2}$ so that Born's rule would give $\langle\mathbf{P}_1\rangle \equiv \langle\psi|\mathbf{P}_1|\psi\rangle = 1/2 = \langle\mathbf{P}_2\rangle \equiv \langle\psi|\mathbf{P}_2|\psi\rangle = 1/2$, obeying the implications of the Probability Symmetry Principle.  However, this choice of the projection operators to depend upon the quantum state violates the Prior Rule Principle.

If instead one follows the PRP and fixes the projection operators first, say as above with fixed normalized complex amplitudes $c_{12}$ and $c_{21}$ and hence fixed $|\psi_{1}\rangle$ and $|\psi_{2}\rangle$, and then allows different possibilities for the quantum state with fixed projection operators, the choice of state $|\psi\rangle = |\psi_{1}\rangle$ gives $\langle\mathbf{P}_1\rangle \equiv \langle\psi|\mathbf{P}_1|\psi\rangle = \langle\psi_1|\psi_{1}\rangle\langle\psi_{1}|\psi_1\rangle = 1 \neq 1/2$ and $\langle\mathbf{P}_2\rangle \equiv \langle\psi|\mathbf{P}_2|\psi\rangle = \langle\psi_1|\psi_{2}\rangle\langle\psi_{2}|\psi_1\rangle = 0 \neq 1/2$.  That is, for an arbitrary quantum state of the assumed form that has observations $O_1$ and $O_2$ definitely occurring each exactly once, Born's rule with the Prior Rule Principle (state allowed to be independent of the projection operators) does not obey the Probability Symmetry Principle (equal probabilities for observations that occur equal numbers of times).  Therefore, if one restricts to theories obeying the Probability Symmetry Principle and the Prior Rule Principle and which have quantum amplitudes for the same observation to occur in different regions, the direct use of Born's rule fails.  Born's rule is not directly applicable in such cosmologies.

\section{Replacing Born's rule by volume averaging}

The preceding Sections show that the direct use of Born's rule with a global quantum state generally fails to give reasonable probabilities for observations in a universe in which the same observation may occur at different locations that the observer cannot distinguish.  Then one cannot find state-independent projection operators $\mathbf{P}_j$ such that the Born-rule observational probabilities $P_j(i) = \langle \mathbf{P}_j \rangle_i$ satisfy reasonable properties such as the Probability Symmetry Principle for a general global quantum state.

Although many replacements of Born's rule are logically possible \cite{cmwvw,insuff,brd,SH}, here I shall give a rather natural procedure for getting the result of what I have called volume averaging \cite{cmwvw,insuff,brd}.  The idea is that although Born's rule does not work for the global quantum state denoted by $\langle\ldots\rangle_i$, it can work indirectly for predicting the ratio of observational probabilities from a local density matrix $\rho(i)$ for a single region, as the ratio of values of $\mathrm{tr}(\mathbf{P}_j\rho(i))$, if one can formulate a procedure for getting such a local density matrix.

Under this class of replacements of the direct Born rule, the logical ambiguity in the replacements of Born's rule would be the ambiguity of how to calculate the local density matrix $\rho(i)$ from the global quantum state.  A particular theory $T_i$ of this form will produce a $\rho(i)$ that depends not only on the global quantum state but also on the procedure for getting $\rho(i)$ from it.  Here I shall outline just one specific procedure for getting a local density matrix $\rho(i)$ whose resulting observational probabilities by the indirect application of Born's rule will obey the Probability Symmetry Principle.

The PSP is not consistent with just taking $\rho(i)$ to be the density matrix of a specific region $L$, if the density matrices of the different regions are different.  Thus one needs some procedure for averaging the density matrices of the different regions.  This is straightforward if there is a definite number of regions $N$, but there is an ambiguity if the quantum state is a superposition of different numbers of regions.  Here I shall make a particular choice within that logical ambiguity, corresponding to what I have elsewhere \cite{cmwvw,insuff,brd} called volume averaging.

Continue with the toy model in which the universe may be divided into a finite number $N$ of different regions (varying with the component of the quantum state), and assume that the quantum state of the universe is the pure state $|\psi\rangle$ with normalized complex coefficients $a_N$ of component states $|\psi_N\rangle$ that each have a definite number $N$ of regions:
\begin{equation}
|\psi\rangle = \sum_{N=1}^{\infty} a_N |\psi_N\rangle, 
\label{state}
\end{equation}
where $\langle\psi_M|\psi_N\rangle = \delta_{MN}$.

Furthermore, write each component state with a definite number $N$ of
regions as a superposition of orthonormal states in the tensor product
of $N$ regions that can each be labeled by either having no observation,
$0$, or by having the observation $j$, in the region $L$, $1 \leq L \leq
N$.  That is, if one lets $\mu_L$ be either $0$ if the region $L$ has no
observer or else $j$ if the region $L$ has the observation $j$, then the
state for a definite $N$ can be written as
\begin{equation}
|\psi_N\rangle = \sum_{\mu_1,\mu_2,\ldots,\mu_N}
 b_{\mu_1\mu_2\ldots\mu_N} |\mu_1\mu_2\ldots\mu_N\rangle, 
\label{N-state}
\end{equation}
where the component state $|\mu_1\mu_2\ldots\mu_N\rangle$ has $\mu_1$
(either no observation, 0, or the observation $O_j$ that is denoted by
$j$) in the first region, $\mu_2$ in the second region, and so on with
all $\mu_L$ for $0\leq L\leq N$ up through $\mu_N$.

So far, I have just set up the notation for the global pure state.  The first step of the procedure to calculate the local density matrix $\rho(i)$ from this global pure state is to construct a global density matrix with no interference terms between states of different $N$,
\begin{equation}
\rho^{\mathrm{global}} = 
\sum_{N=1}^{\infty} |a_N|^2 |\psi_N\rangle\langle\psi_N|.
\label{rho-global}
\end{equation}

The next step of the procedure is to trace over all $N$ regions but $L$ to get a local density matrix for the region $L$ in the case of each fixed $N$:
\begin{equation}
\rho_{LN}(i) = \mathrm{tr}_{K\neq L} (|\psi_N\rangle\langle\psi_N|),
\label{rho-LN}
\end{equation}
Since the observer does not know what region $L$ he or she is in, for each $N$ construct the average density matrix
\begin{equation}
\rho_N(i) = \frac{1}{N}\sum_{L=1}^N\rho_{LN}(i).
\label{rho-N}
\end{equation}
From these averaged density matrices for each region, $\rho_N(i)$, and from the coefficients $|a_N|^2$ of the global density matrix $\rho^{\mathrm{global}}$ corresponding to the entire state (after suppressing the entanglement between different $N$), one can then calculate the final averaged density matrix for a single region:
\begin{equation}
\rho(i) = \sum_{N=1}^{\infty} |a_N|^2 \rho_N(i).
\label{rho}
\end{equation}
Assuming that one starts from a normalized global pure state $|\psi\rangle$, this averaged density matrix $\rho(i)$ for a single region is automatically normalized, $\mathrm{tr}\rho(i) = 1$.

Now one can extract observational probabilities as conditional probabilities using Born's rule on this indirectly constructed averaged density matrix $\rho(i)$.  In particular, let $\mathbf{P}_\mu$ be the projection operator in one region to $\mu$ (either no observation, $\mu = 0$, or the observation $O_j$, $\mu = j$).  Since $\sum_\mu \mathbf{P}_\mu = I$, $\sum_\mu \mathrm{tr}\left(\mathbf{P}_\mu \rho(i)\right) = 1$, so
\begin{equation}
p_\mu(i) = \mathrm{tr}\left(\mathbf{P}_\mu \rho(i)\right)
\label{unnormalized-probabilities}
\end{equation}
can be interpreted as the probability of getting $\mu$ in the normalized density matrix $\rho(i)$, $\sum_\mu p_\mu = 1$.

If the probability calculated this way of getting no observation is positive, $p_0(i) = \mathrm{tr}\left(\mathbf{P}_\mu \rho(i)\right) > 0$, the resulting probabilities of actual observations, $p_j(i)$ with $j>0$, are not normalized to sum to unity over the restricted set of actual observations:  $\sum_{j\geq 0} p_j(i) = 1-p_0(i) < 1$.  Therefore, they cannot be used directly as normalized probabilities of observations.  However, one can readily get normalized probabilities of observations as conditional probabilities, conditionalized upon getting an actual observation ($\mu = j > 0$) rather than no observation ($\mu = 0$):
\begin{equation}
P_j(i) = \frac{p_j(i)}{\sum_{k>0}p_k} = 
\frac{\mathrm{tr}\left(\mathbf{P}_j \rho(i)\right)}
     {\sum_{k>0}\mathrm{tr}\left(\mathbf{P}_k \rho(i)\right)}.
\label{normalized-probabilities}
\end{equation}

One can see that this procedure gives the same result as volume averaging (rather than, say, volume weighting or observational weighting) that I have described elsewhere \cite{cmwvw,insuff,brd}.  It will be left to the reader to give volume weighting or observational weighting in terms of a similar procedure of using Born's rule indirectly with a different averaged density matrix for a single region, but in terms of this type of procedure, those procedures seems somewhat more complicated.  Thus in the method used here, volume averaging seems simpler or more natural than volume weighting or observational weighting.

\section{Conclusions}

The proof above that Born's rule fails in cosmology for a large universe shows that the quantum state of the universe is not enough by itself to give the probabilities of observational results; one needs new rules for extracting these probabilities.  This implies that the measure problem in cosmology (see \cite{cmwvw} for many references) is more serious than might have been thought; it cannot be solved just by knowing the quantum state of the universe.

I have argued previously \cite{cmwvw,insuff,brd} that a partial solution of the measure problem, greatly ameliorating the Boltzmann brain catastrophe, is given by using volume averaging rather than volume weighting.  In this Letter I have shown a simple natural way to get volume averaging by returning to Born's rule indirectly to calculate conditional probabilities of observations from a simple averaged density matrix of a randomly chosen small region of the universe.

\section*{Acknowledgments}

I am grateful for discussions with Andreas Albrecht, Tom Banks, Raphael
Bousso, Sean Carroll, Brandon Carter, Ben Freivogel, Alan Guth, Daniel Harlow, James Hartle, Thomas Hertog, Gary Horowitz, Matthew Kleban, Andrei Linde, Seth Lloyd, Juan Maldacena, Donald Marolf, Mahdiyar Noorbala, Daniel Phillips, Mark Srednicki, Herman Verlinde, Alex Vilenkin, Alexander Westphal, and others.  I have appreciated the hospitality of the Perimeter Institute for Theoretical Physics, where discussions led me to formulate the present prescription for volume averaging.  I am thankful to Stanley Deser and Andreas Albrecht for independently suggesting the personalized title.  This research was supported in part by the Natural Sciences and Engineering Research Council of Canada.


\baselineskip 5pt

\begin{thebibliography}{99}

\bibitem{Born} M.~Born, Z.\ Phys.\ {\bf 37}, 863-867 (1926).

\bibitem{HS} J.~B.~Hartle and M.~Srednicki, Phys.\ Rev.\ D {\bf 75},
123523 (2007) [arXiv:0704.2630].

\bibitem{typdef} D.~N.~Page, ``Typicality Defended,'' arXiv:0707.4169.  

\bibitem{typder} D.~N.~Page, Phys.\ Rev.\ D {\bf 78}, 023514 (2008)
[arXiv:0804.3592].

\bibitem{cmwvw} D.~N.~Page, J.\ Cosmolog.\ Astropart.\ Phys.\ {\bf
0810}, 025 (2008), arXiv:0808.0351.

\bibitem{insuff} D.~N.~Page, Phys.\ Lett.\ B {\bf 678}, 41-44 (2009),
arXiv:0808.0722.

\bibitem{brd} D.~N.~Page, J.\ Cosmolog.\ Astropart.\ Phys.\ {\bf 0708},
008 (2009), arXiv:0903.4888.

\bibitem{SH} M.~Srednicki and J.~B.~Hartle, ``Science in a Very Large Universe,'' arXiv:0906.0042.

\end{thebibliography}
\end{document}